\documentclass{article}
\usepackage{graphicx}
\usepackage{dcolumn}
\usepackage{bm}
\usepackage{psfig}

\begin{document}

\begin{center}
{\bf Opinion Formation on a Deterministic Pseudo-fractal Network}

\bigskip

M.C. Gonz\'{a}lez \footnote{marta@ica1.uni-stuttgart.de},
A.O. Sousa\footnote{sousa@ica1.uni-stuttgart.de},
H.J. Herrmann\footnote{hans@ica1.uni-stuttgart.de}

\bigskip

{Institute for Computer Applications 1 (ICA1), University of 
Stuttgart, Pfaffenwaldring 27, 70569 Stuttgart, Germany.}

\end{center}
\begin{center}
{\bf Abstract}
\end{center}

The Sznajd model of socio-physics, that only a group of people sharing
the same opinion can convince their neighbors, is applied to a scale-free 
random network modeled by a deterministic graph. We also study a model for
elections based on the Sznajd model and the exponent obtained for the
distribution of votes during the transient agrees with those obtained
for real elections in Brazil and India. Our results are compared to 
those obtained using a Barab\'asi-Albert scale-free network.


\section{Introduction}
Growing interest has been focused on the statistical physics of complex 
networks. They describe a wide range of systems in nature and society, 
modeling diverse systems as the Internet, the World Wide Web, the net of 
human sexual contacts, a network of chemicals linked by chemical reactions, 
and social and biological networks \cite{strogatz,Barabasi:RMP,
Barabasi:PhysA,Doro:PhysRevE,doro2,newman}. Such networks exhibit a 
rich set of scaling properties. A number of them are scale-free, that 
is, the probability that a randomly selected node has exactly $k$ links decays 
as a power law, following $P(k)\sim k^{-\gamma}$, where $\gamma$ is the degree
exponent. Consequently they present resilience against random breakdowns, an
effect known as robustness of the network. The shortest-path length
between their sites grows slowly (i.e. logarithmically) with the size of the 
network. That is, compared to the large sizes of networks, the distance 
between their sites are short - a feature known as the ``small world'' effect.

The majority of networks used to generate a scale-free topology are 
stochastic, i.e. the nodes appear to be randomly 
connected to each other. These scale-free random networks have naturally a 
continuous degree distribution. But recently it has been
shown that discrete degree distributions of some deterministic graphs 
asymptotically also exhibit a power law decay \cite{Barabasi:PhysA}. 
Furthermore, scale-free random networks are excellently modeled by such
deterministic graphs\cite{Doro:PhysRevE}. However the comparison 
between the behavior of stochastic and deterministic networks in 
the simulation of a particular model still remains open. 

The involvement of physics in research on spreading of opinion has been 
increasing \cite{stauffer,galam}. Of particular interest here 
is the Sznajd model \cite{sznajd1}, which is one of several recent
consensus-finding models \cite{galam,defuant} and in which each randomly
selected pair of nearest neighbors convinces all its neighbors of their 
opinion, only if the pair shares the same opinion (``Together we stand''); 
otherwise, the neighbor opinions are not affected. One time step means that on
average every lattice node is selected once as the first member of the
pair. It differs from other consensus models by dealing only with 
communication between neighbors, and the information flows outward,
as in rumor spreading. In contrast, in most other models the
information flows inward. Initially, two opinions ($+1$ and $-1$) are
randomly distributed with probability $p$ and $1-p$ respectively over all the 
nodes of the lattice. The basic Sznajd model with random sequential updating 
always leads to a consensus (all sites have the same opinion and the whole
system reaches a fixed point after a certain simulation time). A
phase transition is often observed as a function of the initial
concentration of opinion $p$. A generalization to many different 
opinions (instead of only $\pm 1$) simulated on a Barab\'{a}si-Albert network 
\cite{Barabasi:Science} reproduced quite well the results of the
complex elections of city councillors in the state of Minas Gerais in 
Brazil in 1998 \cite{bernardes}.

In the next section we present the deterministic scale-free network. Then 
we simulate using the Sznajd model an election process on such a network. We 
compare our results with the same simulations carried out on a stochastic 
scale-free network (the Barab\'{a}si-Albert network) and with states 
election's from Brazil and India. 

\begin{figure}[!hbt]
\centering
\includegraphics[width=11.5cm]{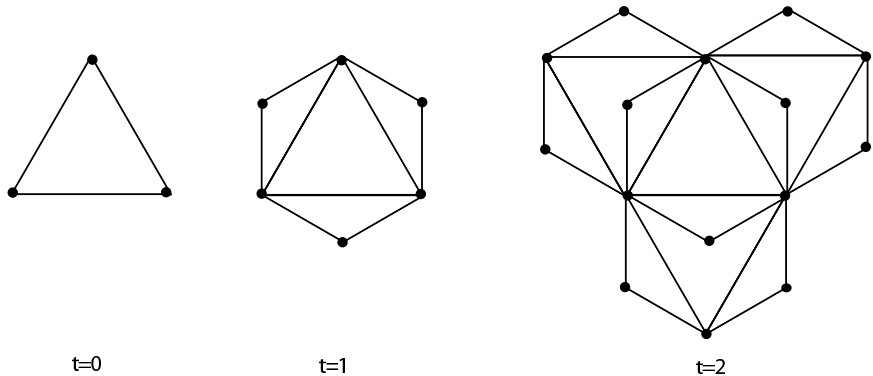}
\caption{The three first generations of the scale-free pseudo-fractal graph. At
each iteration step $t$, every edge generates an additional vertex, which is 
attached to the two vertices of this edge.}
\label{fig:fig1}
\end{figure} 

\section{Models and Results}
The deterministic scale-free graph used in this work grows as follows: At 
each time step, every edge generates an additional vertex, which is attached 
to both end vertices of this edge. Initially, at $t=0$, we have a triangle 
of edges connecting three vertices, at $t=1$, the graph consists of 
$6$ vertices connected by $9$ edges, and so on (see Fig.\ref{fig:fig1}). The 
total number of vertices at iteration $t$ is 
\begin{equation}
N_{t}=\frac{3(3^{t}+1)}{2}
\label{eq:Nt}
\end{equation}
This simple rule produces a complex growing network. Such a graph is called 
a {\it pseudo-fractal}. It has a discrete degree distribution. To relate the 
exponent of this discrete degree distribution to the standard $\gamma$ 
exponent of a continuous degree distribution for random scale-free networks, 
we use a cumulative distribution $P_{cum}(k)$, which follows
\begin{equation}
P_{cum}(k) \sim k^{1-\gamma}
\end{equation}
\noindent
and it is the probability that a node of the network hat at least $k$ 
neighbors. It decreases as a power of $k$ with exponent 
$\gamma = 1+\ln3/\ln2$. In a similar way the average clustering coefficient 
\cite{Doro:PhysRevE}, which is the probability of existence of a link between 
two nodes when they are both neighbors of a same node, can be calculated for 
the infinite graph, $\bar{C} = 4/5$. One obtains a shortest-path-length 
distribution which tends to a Gaussian of width $\sim \sqrt{\ln N}$ centered 
at $\bar{l} \sim \ln N$ for large networks \cite{Doro:PhysRevE}. These 
properties concerning the degree distribution, the clustering coefficient
and the mean length, are also present in a wide range of stochastic scale-free 
networks reported in the literature. They make our simple deterministic 
networks suitable to examine applications very often found on stochastic 
networks. In the next section we apply the Sznajd model on our scale-free 
pseudo-fractal.

\subsection{Monte Carlo simulations of the Sznajd model}

We let the fractal of Fig.\ref{fig:fig1} grow and at each step 
assign sites with random opinion $\pm 1$. At every  
step $t>0$, we have the following process:

\begin{enumerate}
\item The network grows, i.e., $3^{t}$ new sites are added. 

\item A random opinion ($\pm 1$) is set to each new node of the 
network, with probability $p$ ($1-p$) for opinion $+1$ ($-1$).

\item $N_s$ Sznajd runs are performed. For each run, $3^{t}$
sites chosen randomly are analyzed and updated, i.e., one visits for the 
Sznajd model a number of sites equal to the number of sites 
added at that step to the network.
\end{enumerate}

\begin{figure}[!hbt]
\centering
\includegraphics[width=11.5cm]{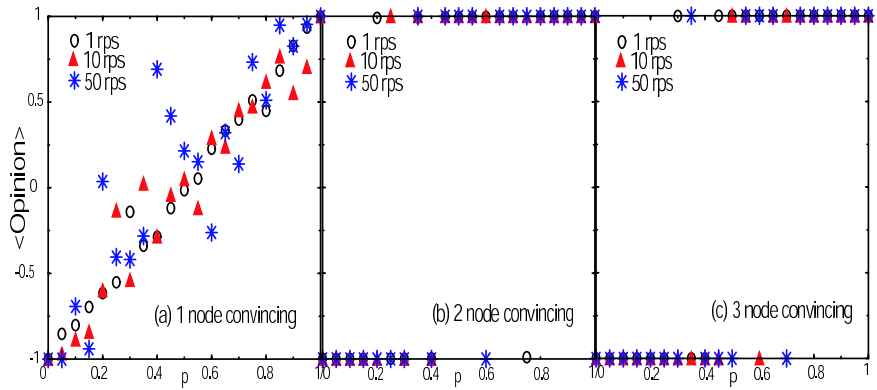}
\caption{Sznajd model on a 29576 nodes pseudo-fractal network with 
$N_{s}=$ 1, 10 and 50 runs per time step for (a) 1 node convincing, 
(b) 2 nodes convincing and (c) 3 nodes convincing.}
\label{fig:fig2}
\end{figure}

Three variations of the Sznajd model on the pseudo-fractal network have been 
investigated:

\begin{itemize}
\item {\bf 1 site convincing:} For each site $i$ chosen, we change the 
opinion of all its neighbors to the site's opinion.

\item {\bf 2 sites convincing:} For each site $i$ chosen, we select randomly 
one of its neighbors. If this selected neighbor has the same opinion of 
the site $i$, then all their neighbors follow the pair's opinion. Otherwise, 
nothing is done.

\item {\bf 3 sites convincing:} For each $i$ site chosen, we select 2 of 
its neighbors at random. If all these three sites have the same opinion, they 
change the opinion of all their neighbors.               
\end{itemize}

\begin{figure}[!hbt]
\centering
\includegraphics[width=11.5cm]{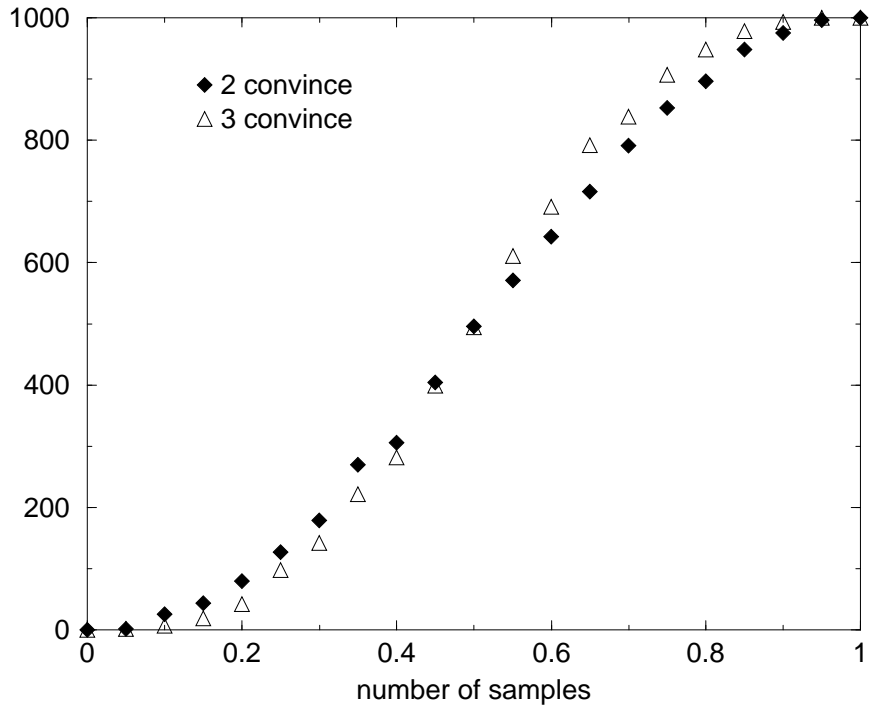}
\caption{Monte Carlo simulation on a 29576 nodes pseudo-fractal network 
counting the number of samples, out of $1000$, for which the fixed point 
all ``up'' is obtained when different values for the initial concentration $p$ 
of nodes ``up'' are simulated. As we can see, the fraction of realizations 
with fixed point up depends on the probability $p$ and on the rule 
implemented, for Sznajd models with 2 and 3 nodes convincing.}
\label{fig:fig3}
\end{figure}

Figure (\ref{fig:fig2}) shows the mean opinion for the three rules 
($1$, $2$ and $3$ sites convincing) for different initial concentrations $p$
and $N_s$ Sznajd runs per step (rps): $1$ rps ($N_{s}=1$) corresponds to one 
realization of Sznajd per step, that is at each step, one chooses $3^{t}$ 
nodes of the network at random to simulate Sznajd. $10$ rps ($N_{s}=10$) and 
$50$ rps ($N_{s}=50$) correspond, respectively, to choose $10 \times 3^{t}$ 
and $50 \times 3^{t}$ nodes randomly per step. We let the network grow up to 
$29576$ sites, that corresponds to $10$ steps. For $1$ node convincing, the 
model can not reach consensus, the results become more unpredictable, the 
larger the number of runs per step (Fig.\ref{fig:fig2}a). However, for two 
and three nodes convincing, a full consensus is observed 
(Figs.\ref{fig:fig2}b and \ref{fig:fig2}c). We clearly see that for rules $2$ 
and $3$ one has a $1st$ order transition since the order parameter (opinion) 
jumps drastically and shows strong hysteresis. Our results are very similar to 
the ones obtained on the Barab\'{a}si-Albert network when the same rules are 
applied \cite{bonnekoh}, except that the latter requires more Sznajd
runs for the network to reach a fixed point, i.e., a full consensus.
  
Since after some time steps the rules $2$ and $3$ always lead to a consensus 
and the whole system reaches a fixed point, in Fig.\ref{fig:fig3} we 
show the number of samples, out of $1000$, for which the fixed point all ``up''
(all with opinion $+1$) is obtained when different values for the initial 
concentration $p$ of nodes ``up'' are simulated. As we can see, the fraction 
of realizations with fixed point ``up'' depends on the probability $p$ and on 
the implemented rule. This is opposed to the results obtained with the same 
rules ($2$ or more nodes convincing) on a square lattice, where an
abrupt change is observed for $p \geq 0.5$ \cite{stauffer2}; but
in agreement with the ones on a Barab\'asi-Abert network \cite{bonnekoh} and 
on a 1-dimensional chain\cite{Sznajd:IntJModPhysCI}. For $1$ node convincing
the system does not tend to a fixed point (consensus), while on square 
lattices it does \cite{Schulze,ochro}, although without the abrupt change of 
2 sites convincing.
 
\begin{figure}[!htb]
\centering
\includegraphics[width=11.5cm]{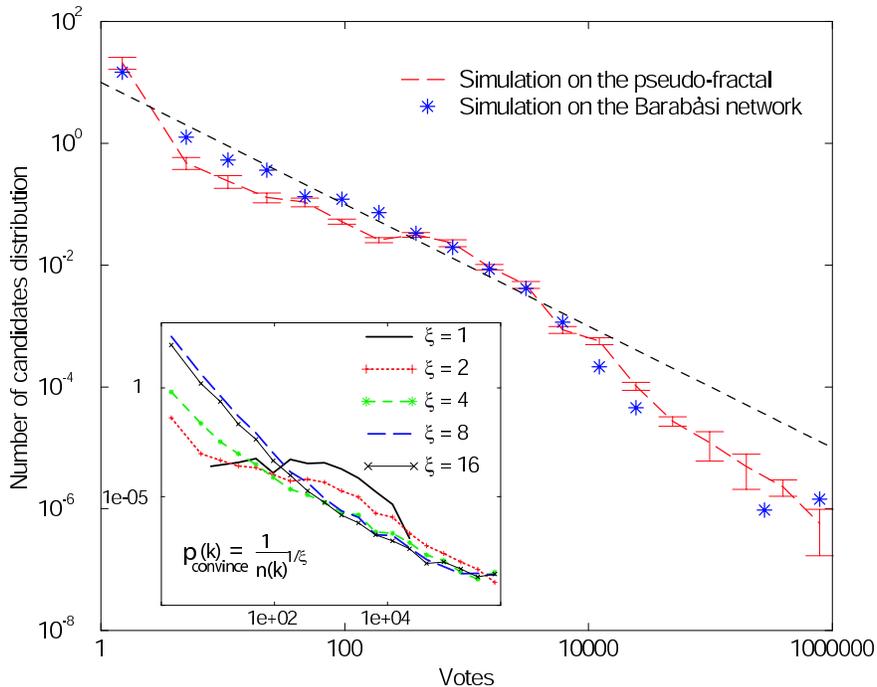}
\caption{Plot of the voting distribution of an election
process on the pseudo-fractal compared to the simulation on a Barab\'asi
network. Both networks have $797163$ sites (voters) and 500 candidates.
We average over $20$ realizations for each network. The distribution for
the pseudo-fractal is after 15 iterations of the convincing process
with $\xi=2$ (Eq. \ref{eq:assumption})
\label{fig:fig4}. The distribution for the Barab\'asi network is after
$83$ iterations. In the inset we show the results of the simulation on the 
pseudo-fractal for $\xi$ $=$ $1$, $2$, $4$, $8$ and $16$, after
$400$, $20$, $4$, $2$ and $1$ iterations respectively. We see that
the results depend on the rule chosen.} 
\end{figure} 

\subsection{Simulation of elections with many candidates}

We create a network of interacting nodes by using the pseudo-fractal network 
prescription as described before (Eq.\ref{eq:Nt}). In addition to our 
network, we also simulated the rule presented in \cite{bernardes} for 
modeling elections on a Barab\'asi network. We summarize the generation of 
the Barab\'asi network as follows: At the beginning of the simulation we have 
$6$ fully connected nodes. After that, we add more and more nodes and connect 
it with $5$ of the present nodes chosen at random. Thus the growth probability 
at any existing node is proportional to the number of nodes already connected 
to it. For comparison we generate both networks with the same number of
nodes (voters), and the same number of candidates.

\begin{figure}[!hbt]
\centering
\includegraphics[width=11.5cm]{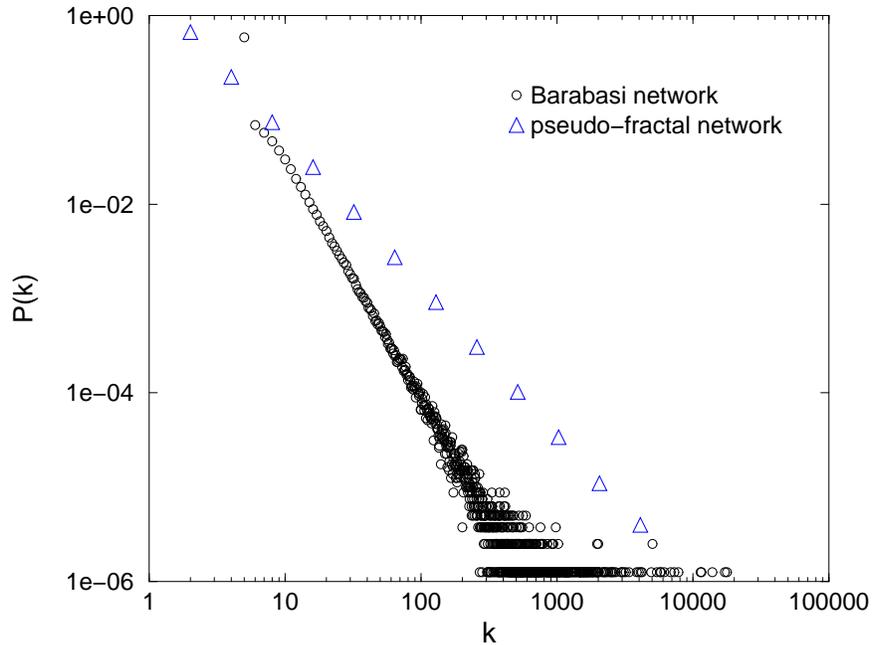}
\caption{Comparison of the degree distributions of the
Barab\'asi network $P(k) \sim k^{-\gamma}$ ($\gamma=2.9$), and of the
pseudo-fractal network, $P(k) \sim k^{1-\gamma}$ ($\gamma=1+\ln3/\ln2$).}
\label{fig:fig5}
\end{figure}

After generating a network, one starts with the ``election process''. $N$ 
candidates are randomly distributed. The value $n$ ($1 \leq n \leq N$) of a 
node on the network represents that this voter has preference
for the candidate $n$. Initially, all sites start with value zero,
meaning that there is no preference for any candidate,
except for $N$ sites that have the number of a candidate. Now, the
electoral campaign starts (only voters with preference for a
candidate can influence other voters, {\it \`a la Sznajd}). At each
time step all the nodes are randomly visited: a random list of nodes
assures that each node is reached exactly once. For each visit, we
implement the following process:

\begin{itemize}

\item We choose a node $i$ at random, if it has preference for a candidate,
we choose among its connected nodes, a node $j$ at random. Otherwise, we 
randomly select another node.

\item If node $j$ has the same candidate as node $i$, each node
convinces all the nodes connected to it with probability:
\begin{equation}
p(k)=\frac{1}{n(k)^{1/\xi}}
\label{eq:assumption}
\end{equation} 
where $n(k)$ is the number of nodes connected to either $i$ or $j$, and 
$\xi>1$ a switching factor that is analyzed later.

\item If node $j$ has no candidate, node $i$ convinces it to
accept its own candidate with the probability $p(i)$ of
eq. (\ref{eq:assumption}).

\item If node $j$ has a different candidate from node $i$, we choose
another node $i$. 
\end{itemize} 

As in real elections, we do not wait for a kind of fixed point
which here would correspond to all the nodes preferring the same candidate,
but we count the votes at an intermediate time. We group in a histogram 
the {\em number of candidates} which received a certain number of {\em votes}. 
Because the bin size for the {\em votes} increases by a factor $2$ for each 
consecutive bin we divide each point of the histogram by the bin size, for 
this reason we have numbers lower than one for the histogram of the {number 
of candidates}, this kind of ``voting distribution'' is used in the literature 
for analyzing similar results \cite{bernardes,Costa:PhysRevE}. In 
Fig.\ref{fig:fig4} we see that the results of the voting distribution for the 
simulations on the pseudo-fractal and on the Barab\'asi network agree very 
well, using $\xi=2$ (Eq. \ref{eq:assumption}) for the pseudo-fractal. In the 
inset we show the results of the simulation on the pseudo-fractal for 
$\xi$ $=$ $1$, $2$, $4$, $8$ and $16$, after $400$, $20$, $4$, $2$ and $1$ 
iterations, respectively. The results differ for each selection of
$\xi$: the system reaches a fixed point more rapidly for larger
parameter $\xi$, i.e, one needs less iterations in order to reach the
same distribution's width. Using the same value used in the
simulations on the Barab\'asi network, $\xi=1$, after nearly $2000$ 
iterations, a fixed point (consensus) on the pseudo-fractal network
could not be reached. A similar result is observed on the square
lattice for the Sznajd model, where if convincing happens only with a certain
probability, then no complete consensus is found \cite{sznajd2,stauffer2}.

\begin{figure}[!hbt]
\centering
\includegraphics[width=8.5cm,angle=-90]{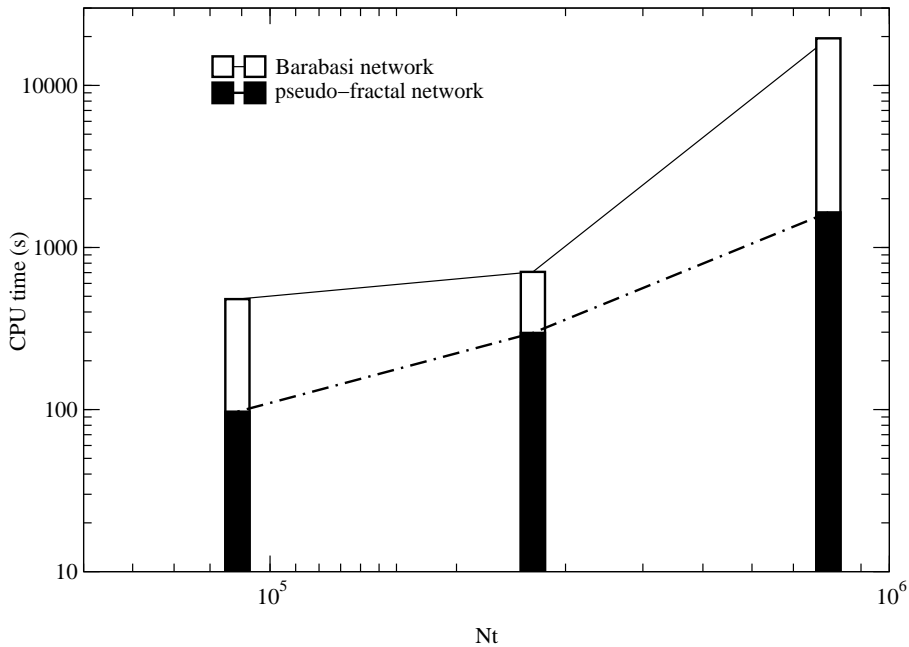}
\caption{CPU time (seconds) on a PIV 2.4 GHz vs. size of the network for the 
pseudo-fractal network and for the Barab\'asi network. We simulate an election 
process with 100 iterations of the convincing model, on networks of sites: 
$88575$, $265722$ and $797163$ with $500$ candidates. The computation time on 
the pseudo-fractal grows linearly with the size of the system, while for the 
stochastic network it grows exponentially.}
\label{fig:fig6}
\end{figure}

\begin{figure}[!hbt]
\centering
\includegraphics[width=11.5cm]{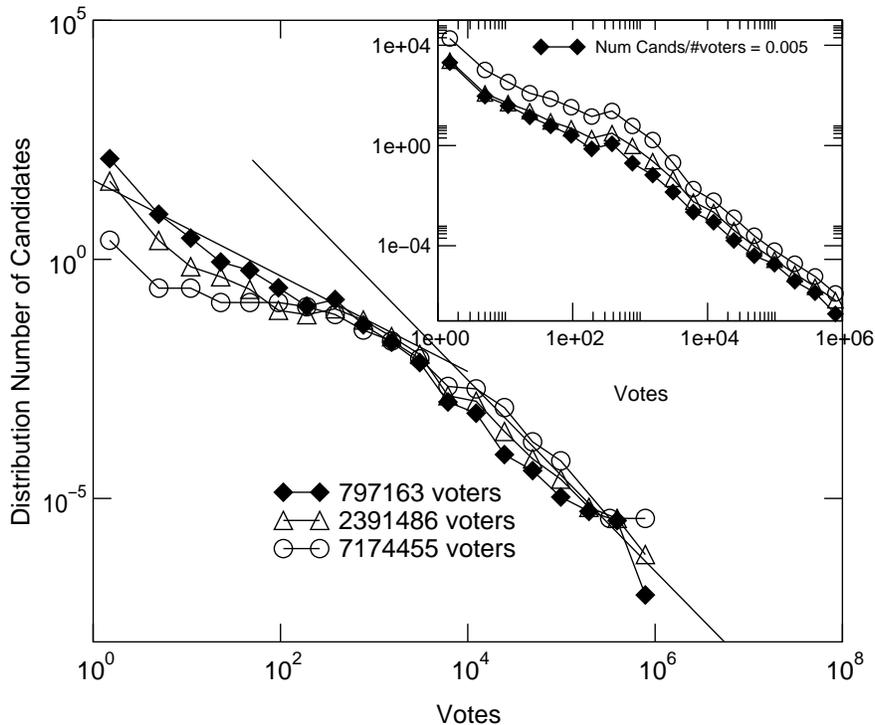}
\caption{Test for size effects on the distribution of candidates for the 
simulation of elections on the pseudo-fractal with 12 generations (diamonds),
13 generations (triangles) and 14 generations (circles) with $2000$ candidates
each, after $15$ iterations of the election process. In the inset we make the 
simulation with a number of candidates proportional to the number of sites of 
the network, keeping the relation $\#Cands/\#voters=0.005$, we see that the 
results are size-independent. The solid lines are guides to the eye with 
slopes $-1$ and $-2$.}
\label{fig:fig7}
\end{figure}

For the simulation on the Barab\'asi network we use $\xi = 1$, like reported
in \cite{bernardes}; this selection means that on average each
node convinces one other node at each process. As we show
in Fig. \ref{fig:fig5} the degree distribution of the pseudo-fractal
is discrete and is given by $P_{cum}(k)=k^{-\ln3/\ln2}$, while
the degree exponent for the Barab\'asi network presented in this
work is $\gamma=2.9$ \cite{Barabasi:Science}. Knowing $P(k)$ we can calculate
the probability distribution of convincing choosing one site at random,
that is the probability of convincing Eq.\ref{eq:assumption}
multiplied by the degree distribution $P(k)$.
For the convincing distribution of both networks to be similar,
one has to chose Eq. \ref{eq:assumption} with $\xi=2$ for the simulation
on the pseudo-fractal. We obtain qualitatively
the same results in both simulations, being $12$ time
faster in the pseudo-fractal than in the Barab\'asi network, because 
it requires less memory space and computation 
time (see Fig.\ref{fig:fig6}). 

In Fig. \ref{fig:fig7} we see that the shape of the voting distribution
after a given number of iterations does not change with the size of the 
network. Changing the number of candidates doesn't alter the 
form of the distribution (see the inset of Fig. \ref{fig:fig7}).     

\begin{figure}[!hbt]
\centering
\includegraphics[width=11.5cm]{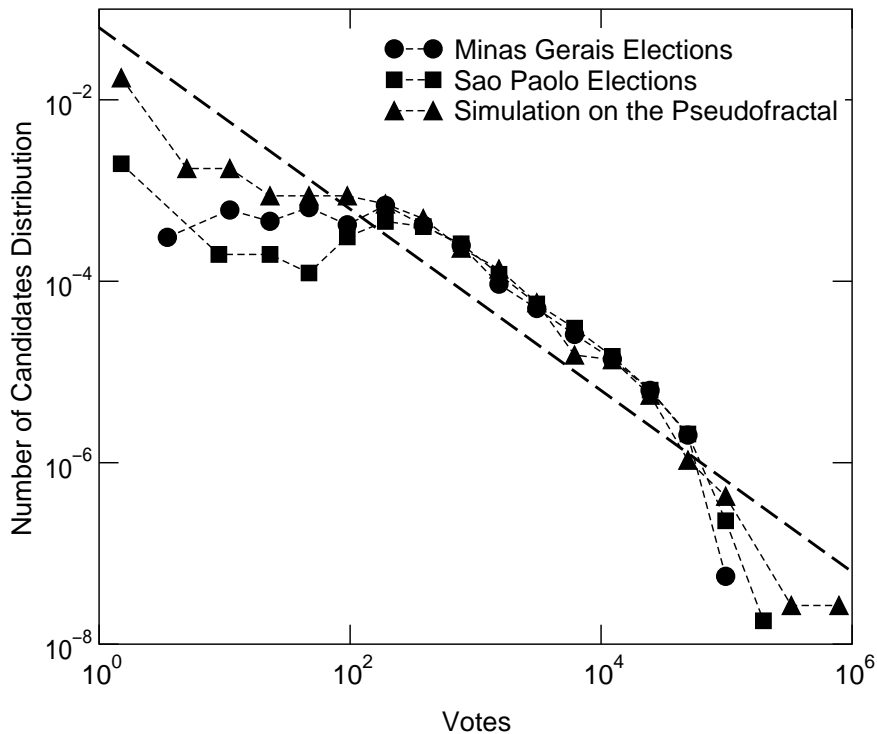}
\caption{Result of the simulation of an election process after 20 iterations 
on a pseudo-fractal network of $21 523 362$ nodes and $1144$ candidates 
(triangles). Compared to the voting distribution for the state of Minas Gerais 
in 1998 (circles) ($819$ candidates, $11 815 183$ voters) and the state of 
S\~ao Paulo (1998) (squares) ($1260$ candidates, $23 321 034$ voters). Both 
axis are plotted on a logarithmic scale. The dashed  straight line is a guide 
to the eye with slope $-1$. The bin size for the $votes$ increases by a factor 
$2$ for each consecutive bin. The height of the distribution of the 
pseudo-fractal is multiplied by 10 to see better the comparison of the 
results.}
\label{fig:fig8}
\end{figure}

We consider the results from two Brazilian states (S\~ao Paulo and Minas 
Gerais) for the positions of local state deputies. In such elections the 
voters vote directly for the candidate and not for parties. Some elections 
occur with a high number of candidates, of the order of thousands, and with 
a number of voters of the order of millions. The official results for each 
state are available on the Internet \cite{brazil}.

In Fig.\ref{fig:fig8} we see the results of a simulation on a $15$-generations 
pseudo-fractal ($21523362$ nodes) and $1144$ candidates, and compare it to 
the results of real elections described in Fig.\ref{fig:fig8}. For the moment 
we do not take into account abstentions, or invalid votes in the simulation.
The pseudo-fractal tends to consensus in a similar way as the real elections. 
After averaging over more than $100$ realizations we see, that the deviations 
from a perfect line in the intermediate regions of votes, seen in the real 
elections, are not of statistical nature but seem due to the determinism of 
the network. The general behavior of the distribution of candidates of the 
simulation of the elections on the pseudo-fractal however agrees with the one 
observed in real results. The results of the distribution  of candidates for 
the simulated and the real cases follow hyperbolic law,
\begin{equation}
   N(v)\propto 1/v,
\label{eq:hyperbolic}
\end{equation}     
for the number $N$ of candidates having $v$ votes, 
extending over two or three order of magnitude, with deviations only
for small and large numbers of votes \cite{Costa:PhysRevE}.

\begin{figure}[!ht]
\centering
\includegraphics[width=11.5cm]{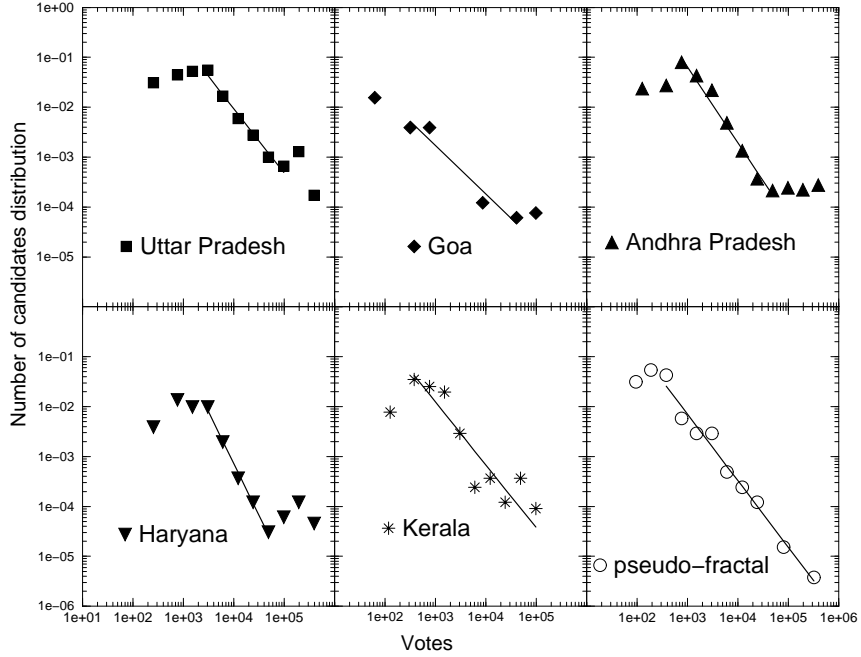}
\caption{Voting distribution for state elections of India in 1998. For Uttar 
Pradesh (squares) with $55,015,804$ voters and $649$ candidates. Goa 
(diamonds) with $532,766$ voters and 12 candidates. Andhra Pradesh (triangles 
up) with $31,829,338$ voters and $301$ candidates. Haryana (triangles down) 
with $7 516,884$ votes and 84 candidates and Kerala (stars) with $13,036,581$ 
voters and $108$ candidates. We obtain results qualitatively comparable to 
this kind of processes with a Pseudo-fractal with 12 generations and $80$ 
candidates, after few iterations ($5$). The solid lines are squares fits to 
the data, in the intermediate regions. The slopes are: $-1.32$ (Uttar 
Pradesh), $-0.97$ (Goa), $-1.51$ (Andhra Pradesh), $-2.06$ (Haryana), $-1.26$ 
(Kerala) and $-1.32$ (Pseudo-fractal)}
\label{fig:fig9}
\end{figure}

In addition we analyzed the behavior of the voting distribution for the 
elections in India to the lower house of the Parliament (Lok Sabha). These 
elections are events involving political mobility and organizational mobility 
on an amazing scale. In the 1998 election to Lok Sabha there were $1269$ 
candidates from $38$ officially recognized national and state parties seeking 
election,$1048$ candidates from registered parties not recognized and $10635$ 
independent candidates. A total number of $596185335$ people voted. The 
Election Commission employed almost $400000$ people to run the election. The 
official results of this elections are available on the Internet \cite{india}. 
In contrast to the Brazilian elections, in this process the country is 
divided into $543$ parliamentary constituencies, each of which return one 
representative to the Lok Sabha. That is, not all the voters elect among all 
the candidates of one state (like in Brazil), but there is one election 
process for each of the $543$ representatives, that occurs in each 
parliamentary constituency. These parliamentary constituencies are selected 
by an independent Delimitation Commission, which creates constituencies which 
have roughly the same population, subject to geographical considerations and 
the boundary of the states and administrative areas.

\begin{figure}[!hbt]
\centering
\includegraphics[height=9cm]{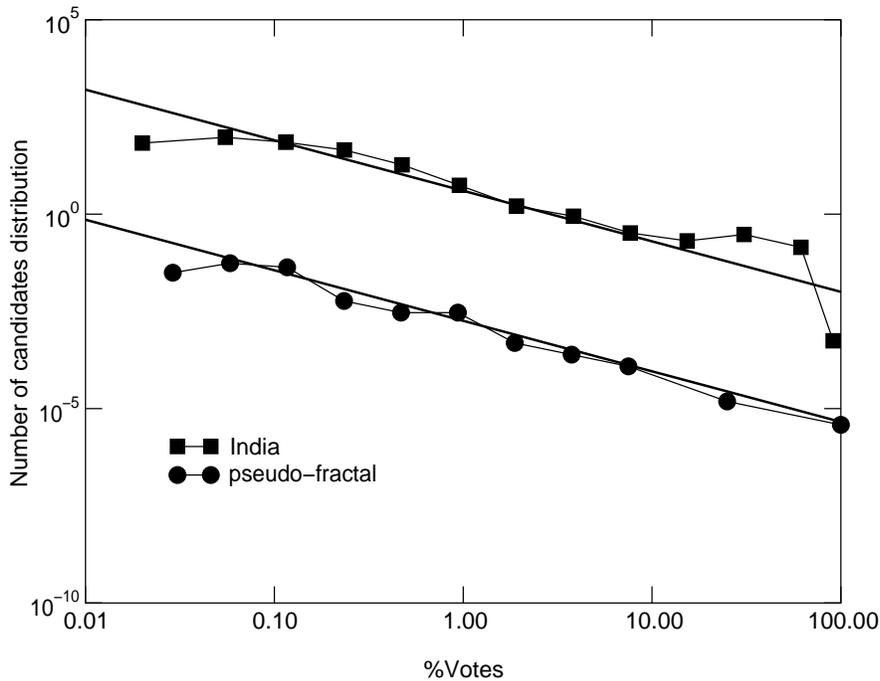}
\caption{Voting distribution for India in 1998 (squares),
compared to the simulation on a Pseudo-fractal (circles) 
with 12 generations and $80$ candidates, after few iterations ($5$).
The solid lines are guide to the eye with slope $-1.3$}
\label{fig:fig10}
\end{figure}

At the end of the process one can analyze the voting distribution of each 
state of the India. In Fig. \ref{fig:fig9} we observe the results for 5 
states: Uttar Pradesh, Goa, Andhra Pradesh, Haryana and Kerala, with 85, 2, 
42, 10 and 20 constituencies respectively. The voting distribution for each 
state is the superposition of different electoral processes for all the 
constituencies of the state. For each election in a constituency there is 
a voting result that corresponds to few candidates,  between $5$ and $10$.
The difference between the India and Brazilian results appears at
this stage. If one analyses the election process of each parliamentary
constituency, even though one does not have a large number of
candidates for the statistics, a distribution with slope $\sim -1$ is 
observed. However, the final results of the process for each Indian state
provide a different profile far from the hyperbolic one ($1/v$ type of
distribution) (see Fig.\ref{fig:fig9}), which was observed for many 
Brazilian states \cite{bernardes,Costa:PhysRevE}. In order to
analyze the number of candidates $N$ which received a certain fraction
of votes $v$ for the nationwide voting process, we have normalized the votes
of each candidate by the total number of voters (Fig.\ref{fig:fig10}). As
can be seen, the number of candidates $\rm N$ follows a power law 
$N(v) \propto v^{\alpha}$ , with $\alpha=-1.3$ (for Brazil,
$\alpha=-1$). 

The differences in the Sznajd simulations on the pseudo-fractal network
for the India and Brazilian elections are mainly due to the number of 
candidates considered for each one. In the latter, the number of candidates is
almost $0.005\%$ of the lattice nodes, while for the former it is
$0.01\%$. The fixed point (consensus) is reached the faster, the larger
the density of candidates. Because of that, the comparison of the real
elections in India with our simulations were made taking account only
$5$ iterations of the Sznajd model on the pseudo-fractal network ($20$
iterations for Brazilian elections). 

\section{Conclusions}
We studied the behavior of a deterministic scale-free network simulating
a spreading of opinion model on it. We solved the Sznajd's model on a growing
deterministic scale-free network, obtaining always a consensus of opinion 
after some runs of the model. The final opinion presents a $1st$ order 
transition, that shows strong hysteresis, as a function of the initial 
concentration of opinions in the network for rules $2$ and $3$. The results 
coincide with the results reported for a stochastic scale free network 
\cite{bonnekoh}. We found that there is not scaling in the finite-size cut-off
showing that the system is not critical (Fig.\ref{fig:fig7}).
We simulated election processes on the network. The probability of
convincing of the model has to be adapted ($\xi=2$, in Eq. \ref{eq:assumption})
in order to avoid the differences in the values of the exponents $\gamma$ 
of the degree distribution of our deterministic network and the $\gamma$ 
exponent of a typical stochastic network (Fig. \ref{fig:fig4}). We obtained 
the same results, reported for a stochastic scale-free network, but with 
computation times considerably lower (see Fig.\ref{fig:fig6}). We could 
use our model to reproduce qualitatively good complex electoral processes,
such as the states elections for parliament in Brazil and India. 

\medskip
\noindent{\bf Acknowledgments}

\noindent We thank D. Stauffer for helpful discussions and a critical 
reading of the manu\-script; A.O. Sousa acknowledges a research grant from 
the Alexander von Humboldt Foundation and M.C. Gonz\'{a}lez a fellowship 
from DAAD.

\end{document}